# OpenCitations, an open e-infrastructure to foster maximum reuse of citation data

Chiara Di Giambattista
Department of Classical Philology and Italian Studies, University of Bologna, Bologna, Italy
Communications Director and Community Development Manager of OpenCitations

Ivan Heibi
Department of Classical Philology and Italian Studies, University of Bologna, Bologna, Italy
Responsible for the OpenCitations' technical infrastructure

Silvio Peroni
Department of Classical Philology and Italian Studies, University of Bologna, Bologna, Italy
Director of OpenCitations

David Shotton
Oxford e-Research Centre, University of Oxford, Oxford, United Kingdom
Director of OpenCitations

## Abstract

OpenCitations is an independent not-for-profit infrastructure organization for open scholarship dedicated to the publication of open bibliographic and citation data by the use of Semantic Web (Linked Data) technologies. OpenCitations collaborates with projects that are part of the Open Science ecosystem and complies with the UNESCO founding principles of Open Science, the I4OC recommendations, and the FAIR data principles that data should be Findable, Accessible, Interoperable and Reusable. Since its data satisfies all the Reuse guidelines provided by FAIR in terms of richness, provenance, usage licenses and domain-relevant community standards, OpenCitations provides an example of a successful open e-infrastructure in which the reusability of data is integral to its mission.

Correspondence should be addressed to Ivan Heibi. Email: ivan.heibi2@unibo.it

OpenCitations (http://opencitations.net ) is an independent not-for-profit infrastructure organization for open scholarship dedicated to the publication of open bibliographic and citation data by the use of Semantic Web (Linked Data) technologies (Silvio Peroni and David Shotton, 2020). It also develops and shares open-source software and open services. Bibliographic citation data have a fundamental value for the world of scholarship, and their open availability is a crucial requirement for the bibliometrics and scientometrics domain. OpenCitations thus assists the creation of reproducible metrics for research assessment exercises. It is also engaged in advocacy for open citations, particularly in its role as a key founding member of the Initiative for Open Citations (I4OC). OpenCitations fully espouses the UNESCO founding principles of Open Science, complies with the recommendations of I4OC that citation data should be structured, separable, and open, and has fully adopted the Force11 FAIR data principles that data should be findable, accessible, interoperable and re-usable.

OpenCitations collaborates with projects that are part of the Open Science ecosystem and that share similar values. The FAIR data principles, in particular, are of central relevance to data reuse within the scope of the EC-funded project OpenAIRE, which advances Open Science and provides a platform whereby a large-scale collection of research outputs is made easily discoverable and re-usable. The OpenAIRE-Nexus project started in January 2021 to embrace and expand the operation of a portfolio of thirteen services, provided by OpenAIRE infrastructure, public institutions, organizations and universities, which it classifies into three portfolios entitled PUBLISH, MONITOR, and DISCOVER. The OpenAIRE Nexus portfolios thus focus on the demands of the three main categories of the research lifecycle, and make sure that these services are integrated to provide a uniform Open Science Scholarly Communication package for the European Open Science Cloud (EOSC). OpenCitations is involved in the OpenAIRE-Nexus project by providig open bibliographic citations. By integrating its functionalities with the OpenAIRE Research Graph and other OpenAIRE-Nexus services within the EOSC Resource Catalogue, OpenCitations participates in the creation of an e-infrastructure that enables the online publication of research "for the benefit of the open science community worldwide".
Since, by definition of the UNESCO Recommendations of Open Science, science practice is "open" if it makes "scientific knowledge openly available, accessible and reusable for everyone", re-usability is one of the main foci of OpenCitations, whose citation data satisfies all the Reuse guidelines provided by FAIR in terms of richness, provenance, usage licenses and domain-relevant community standards.

## Data richness

OpenCitations provides information on scholarly citations between publications covering all academic subject domains. OpenCitations' main database, the OpenCitations Index of Crossref open DOI-to-DOI citations (COCI) (Ivan Heibi et al., 2019), now contains more than 1.29 billion citation links (as of March 2022) from more than 72 million bibliographic resources. According to a recent independent analysis by Alberto Martìn-Martìn (Alberto Martín-Martín, 2021), OpenCitations' coverage is now approaching parity with that of the two principle proprietary citation indexes, Web of Science and Scopus, as shown in Figure 1:

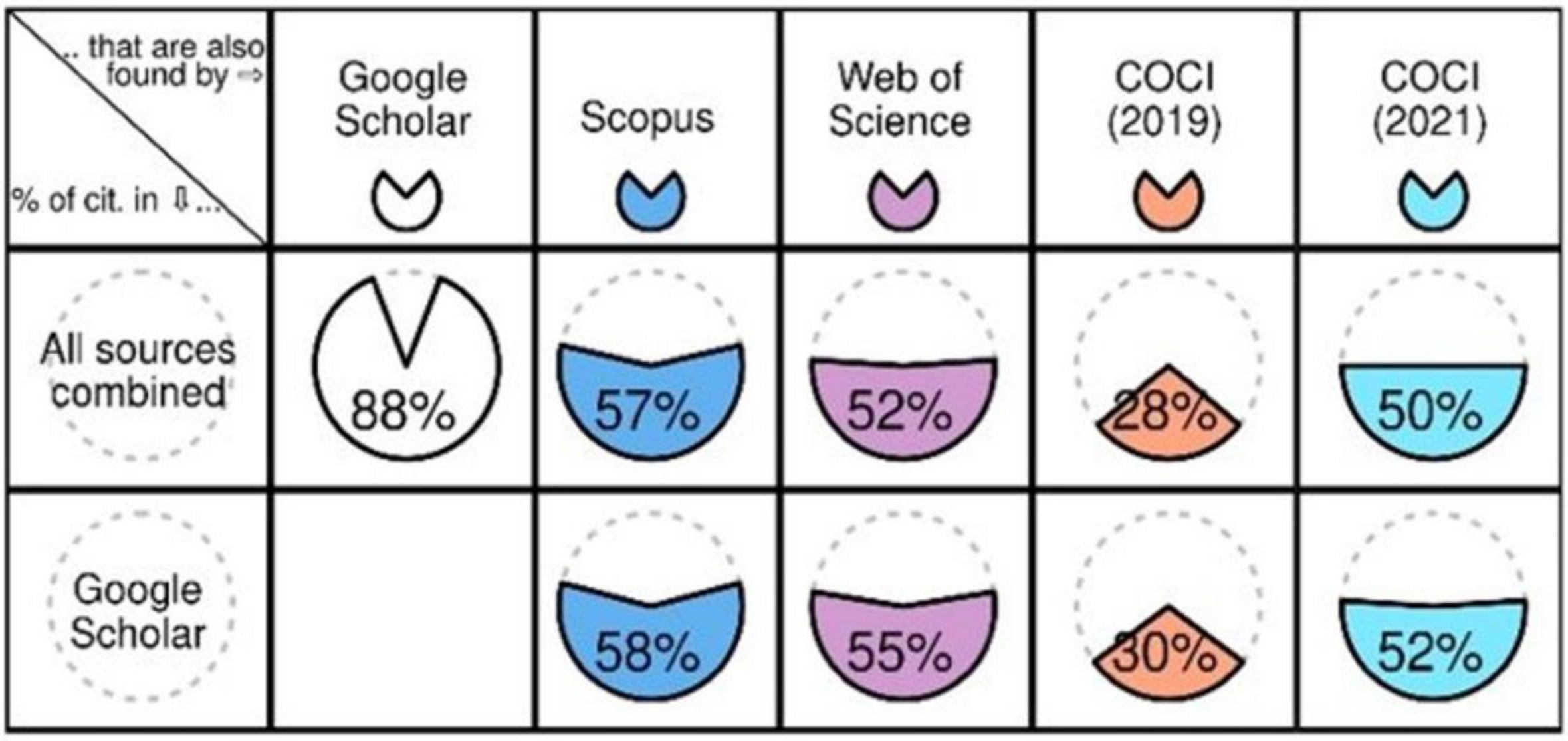

**Figure 1.** Percentage of citations found by each database, relative to all citations (first row), and relative to the number of citations found by a specific database (Google Scholar). From Alberto Martín-Martín (2021)

## Provenance information

Every item of data within OpenCitations is accompanied by machine-readable metadata recording the provenance of the information, including its source, the nature of any subsequent curatorial modification or correction, the dates of such actions, and the identities of the agents involved in obtaining or changing it (whether human or computational).

## Usage licenses

To ensure the greatest possible reusability, all OpenCitations data are published under a Creative Commons CC0 Public Domain Waiver that permits downloading and re-use of any nature, including commercial. Thus existing non-for-profit and commercial providers of citation indexes may themselves use OpenCitations data to expand their coverage and to enhance the value of their charged-for added-value services build over their data holdings.

## Domain-relevant community standards

OpenCitations activities adopt Semantic Web technologies for data description, particularly the Resource Description Framework (RDF) and the Web Ontology Language (OWL). Moreover, OpenCitations treats citations as first class data entities, each with a unique persistent identifier, an Open Citation Identifier (OCI), rather than simply as relationships between citing and cited publications. In this way, OpenCitations is able to assign descriptive metadata to each citation individually, for example its citation timespan, and whether or not it is a self-citation (author self-citation, journal self-citation, institutional self-citation, etc.), encoded using CiTO, the citation typing ontology. The OpenCitations Data Model (Marilena Daquino et al., 2020) is used to structure all OpenCitations RDF data, facilitating the semantic interoperability with related data from other providers of open scholarly information (Figure 2).

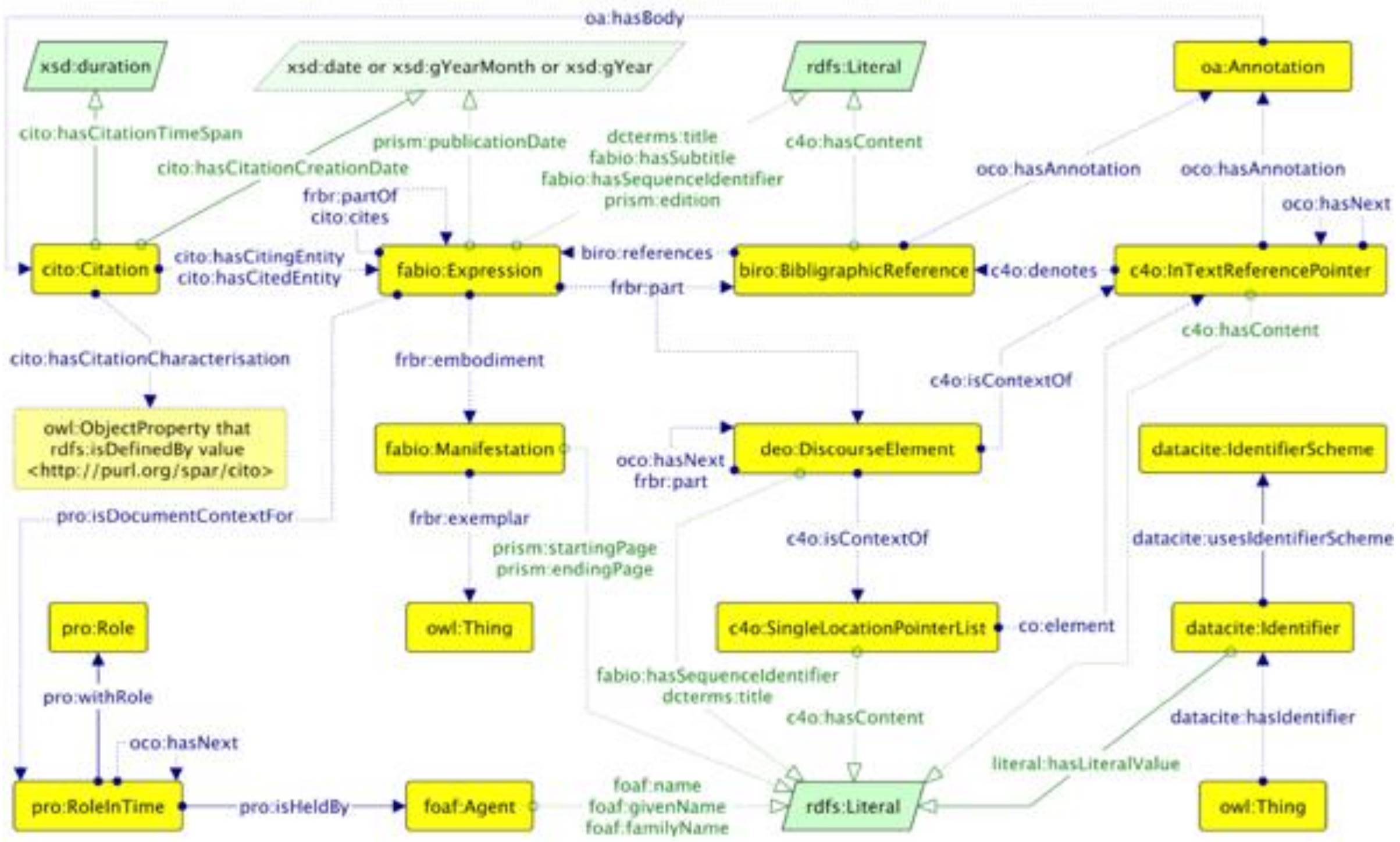

**Figure 2.** The Graffoo diagram of the main ontological entities described in the OpenCitations Data Model.

In addition, our open source software, used to run all our databases and services, not only permits third parties to re-use our developments but also ensures that the OpenCitations system can be re-created elsewhere in the unlikely event that OpenCitations itself ceases to operate, thereby ensuring the preservation and longevity of our work and data.

In conclusion, OpenCitations provides an example of a successful open e-infrastructure in which the reusability of data is integral to its mission, namely “to make global scholarly citation data available at zero cost and without restriction for third party analysis and reuse, in both human- and machine-readable formats”.

# References

Marilena Daquino, Silvio Peroni, David Shotton, Giovanni Colavizza, Benham Ghavimi, Anne Lauscher, Philipp Mayr, Matteo Romanello, Philipp Zumstein (2020). The OpenCitations Data Model. In Proceedings of the 20th International Semantic Web Conference (ISWC 2020). https://doi.org/10.1007/978-3-030-62466-8_28.

Alberto Martín-Martín (2021). “Coverage of open citation data approaches parity with Web of Science and Scopus”, OpenCitations blog, October 27, 2021: https://opencitations.wordpress.com/2021/10/27/coverage-of-open-citation-data-approaches-parity-with-web-of-science-and-scopus/

Ivan Heibi, Silvio Peroni, David Shotton (2019). Software review: COCI, the OpenCitations Index of Crossref open DOI-to-DOI citations. Scientometrics, 121 (2): 1213-1228. https://doi.org/10.1007/s11192-019-03217-6

Silvio Peroni, David Shotton (2020). OpenCitations, an infrastructure organization for open scholarship. Quantitative Science Studies, 1(1): 428-444. https://doi.org/10.1162/qss_a_00023